# Point-focus spectral splitting solar concentrator for multiple cells concentrating photovoltaic system

Carlo Maragliano[1*], Matteo Chiesa[1] and Marco Stefancich[2]

[1] Laboratory for Energy and NanoScience (LENS), Institute Center for Future Energy Systems (iFES), Masdar Institute of Science and Technology, P.O. Box 54224, Abu Dhabi, UAE.
[2] Istituto Materiali per l'Elettronica ed il Magnetismo, Consiglio Nazionale delle Ricerche, Parco Area delle Scienze 37/A - 43124 Parma, Italy
[*] cmaragliano@masdar.ac.ae

**Abstract:** In this paper we present and experimentally validate a low-cost design of a spectral splitting concentrator for the efficient conversion of solar energy. The optical device consists of a dispersive prismatic lens made of polycarbonate designed to simultaneously concentrate the solar light and split it into its spectral components. With respect to our previous implementation, this device concentrates the light along two axes and generates a light pattern compatible with the dimensions of a set of concentrating photovoltaic cells while providing a higher concentration ratio. The mathematical framework and the constructive approach used for the design are presented and the device performance is simulated using a ray-tracing software. We obtain spectral separation in the visible range within a 3x1 $cm^2$ area and a maximum concentration of 210x for a single wavelength. The device is fabricated by injection molding and its performance is experimentally investigated. We measure an optical transmissivity above 90% in the range 400-800 nm and we observe a spectral distribution in good accordance with simulations. Our results demonstrate the feasibility of the device for cost effective high efficiency concentrated photovoltaic systems.

## 1. Introduction

Solar energy holds the potential to become a cost-effective source of energy in the near future[1]. Two are the main pathways for harvesting solar energy: solar thermal and photovoltaic. Solar thermal collectors transform the solar radiation into heat that can be either stored for later use or directly fed to further conversion stages, while photovoltaic receivers (PV cells) can produce electricity directly from sunlight. Differently from thermal absorbers, which are able to capture the entire solar spectrum, PV cells have a fixed spectral response that depends intrinsically on the nature of the material. Semiconducting materials absorb photons having energies greater than the band gap: the band gap energy is partially transferred to useful electrical charge carriers, while the excess energy is quickly dissipated as heat after carriers' thermalization[2, 3]. On the other hand, photons with energies lower than the material band gap do not generate carriers and are often weakly absorbed. Given the wide spectral content of the sunlight (most of the power lies between 350 and 2000 nm, corresponding to a range of energies from 3.55 to 0.62 eV), a single material cannot efficiently generate electricity. Shockley and Queisser studied the theoretical limit of the efficiency of a single junction solar cell and found that the limit for a silicon PV cell under 1 sun illumination is approximately 30%[4]. Polman and Atwater calculated the energy losses due to thermalization and lack of absorption in a single junction solar cell and concluded that these two mechanisms account for at least 40-45% of the total losses[5]. Reducing these losses, even partially, could allow the development of solar cells with efficiencies above 50%.

An effective approach to alleviate the spectral mismatch issue of solar cells consists in simultaneously using different absorbers to harvest the solar energy. Such approach has brought to the development of multi-junction (MJ) solar cells, where cells with decreasing band gaps are stacked on top of each other and connected in series[6]. This design allows the top layer to absorb the high-energy photons, while the lower layers absorb the lower energy photons that cannot be absorbed by the top layer. Although the effectiveness of such approach has led to commercial cells exceeding 40% of efficiency and laboratory efficiencies as high as 44.7 %[7], the stacking of different mono-crystalline materials requires a tight lattice matching, thus imposing a severe constraint on the materials choice. Sub cells current matching requirements due to the series connection and the need for tunnel junctions between sub-cells further limit the design freedom. The resulting fabrication process of MJ cells, generally based on stacked mono-crystalline III-V materials, requires expensive epitaxial growth techniques and is, for this reason, economically viable for terrestrial application only under very high concentration (around or above 500x)[6]. The high optical concentration level, in turn, raises a cascade of optical design issues and leads to very small acceptance angles of the system, thus imposing demanding and expensive requirements on the mechanical tracking system[8]. Although possible solutions to the drawbacks of MJ cells have been proposed in the literature[9-11], such approach still remains limited to niche applications.

An alternative approach to MJ solar cells consists in separating the sunlight into various wavelength bands and directing each band to band-matched absorbers[12, 13]. With respect to MJ solar cells, a spectral splitting solar system does not impose any limit on the choice of materials or manufacturing process, as the cells are physically independent. Cells are arranged in tandem on the same supporting structure and can be therefore independently manufactured with any technique, without concern for lattice mismatch. Moreover, current matching among the different absorbers is not necessarily required as separated converters can be used for each homogeneous cell group or different multi-cell parallel/series connections can be devised. Finally, spectral splitter systems, allowing for cheaper cells, open the way to lower concentration and higher acceptance angle optics with respect to MJ solar cell systems, leading to more relaxed requirements on optics and mechanical solar tracking of the setup.

Different designs have been proposed in the literature to achieve spectrum separation. Spectrum splitting holographic elements, which are fabricated recording an interference pattern on a holographic medium, represent a relatively cheap solution for spectral splitting applications [14-16], but suffer from low optical efficiencies due to noise or undesired interference effects [13]. Thin diffractive-optical elements, which are characterized by a micro-structured surface, have been used for spectrum splitting applications[17-20] because of their versatility, which allows achieving different spectral combinations. Nevertheless, the high cost and complexity has limited their use. Dichroic filters, made of periodic sequences of thin dielectric non-absorbing layers with high refractive index contrast, represent an efficient solution for spectral splitting applications[21, 22], but require expensive deposition techniques as well as precise control of layer thicknesses. The same applies to rugate filters, which are generally more robust and have a higher durability with respect to thin-film optical filters, but require expensive fabrication techniques and are highly sensitive to light incidence angle[23].

In this paper we present a low-cost design of a dispersive optical element capable of simultaneously concentrating and splitting the solar light. The element consists of a sequence of trapezoidal prims arranged in a curved shape envelope. Each prism works independently and their orientation is designed in a way that the spectrum generated by each of them sums up at the focal plane: this allows obtaining spectral separation while increasing the light concentration. The element is designed using polycarbonate (PC), a transparent plastic material, but the design procedure can be applied to any optically dispersive material. With respect to our previous implementation[24], this element concentrates the incoming light along two axes on a finite area matching the dimensions of multiple concentrating photovoltaic cells. The device is fabricated by injection molding, which allows for low-cost mass scale production. In the following sections we present the conceptual framework and the mathematical building model used to design the element. The performance of the element is simulated using ray-tracing software and the validity of the design is verified experimentally on a prototype. Our characterization aims at proving the feasibility of the element for cost effective high efficiency concentrated photovoltaic systems.

2. **Optical design principles**

The optical design follows the same principles described in [24] where we presented an optical element defined as a set of solid transparent components, each operating independently. The basic component is a simple dispersive triangular prism. In optics, dispersive prisms are used to split a white light beam into its different colours: the separation among distinct wavelengths depends on the geometry of the prism and on the inherent dispersive characteristic of the material it is constructed of. Figure 1a shows a schematic of the splitting by a prism. For simplicity, we define the optical z-axis to be parallel to the direction of the light incident on the prism, assumed to be perfectly collimated. The geometrical parameters that affect the prism behaviour are the angle formed by the normal of the entrance facet with the light rays (defined as $I_1$) and the apical angle of the prism (defined as $A$). The material parameter is the index of refraction $n_\lambda$. Based on the parameters $I_1$ and $A$, we derived an analytical formula for the deviation angle between the incoming and the exiting ray $D_\lambda$ [24]:

$$D_\lambda = I_1 - A + \sin^{-1}\left\{n_\lambda \cdot \sin\left[A - \sin^{-1}\left(\frac{1}{n_\lambda} \cdot \sin(I_1)\right)\right]\right\} \qquad (1)$$

Equation (1) shows that for a given value of $I_1$ and $A$, the deviation angle depends uniquely on the refractive index of the material $n_\lambda$. Given that $n_\lambda$ depends on the light wavelength according to the dispersion characteristics of the material, different colors will be refracted along separate angular directions. The result of the separation can be appreciated observing the light pattern on a receiver positioned away from the prism.

The design presented in [24] combines the splitting effect of many prisms to allow also for light concentration. Figure 1b shows the concept of the prismatic element. A set of prisms is arranged along a curved line in a way that the light rays of a specific wavelength converge in the same point at a distance Z from the element. The superposition of each prism contribution results then in a concentrated and spectrally divided beam. The distance Z is mainly determined by the curvature of the lens, while the concentration ratio depends on the number of prisms: the greater the number, the greater the concentration.

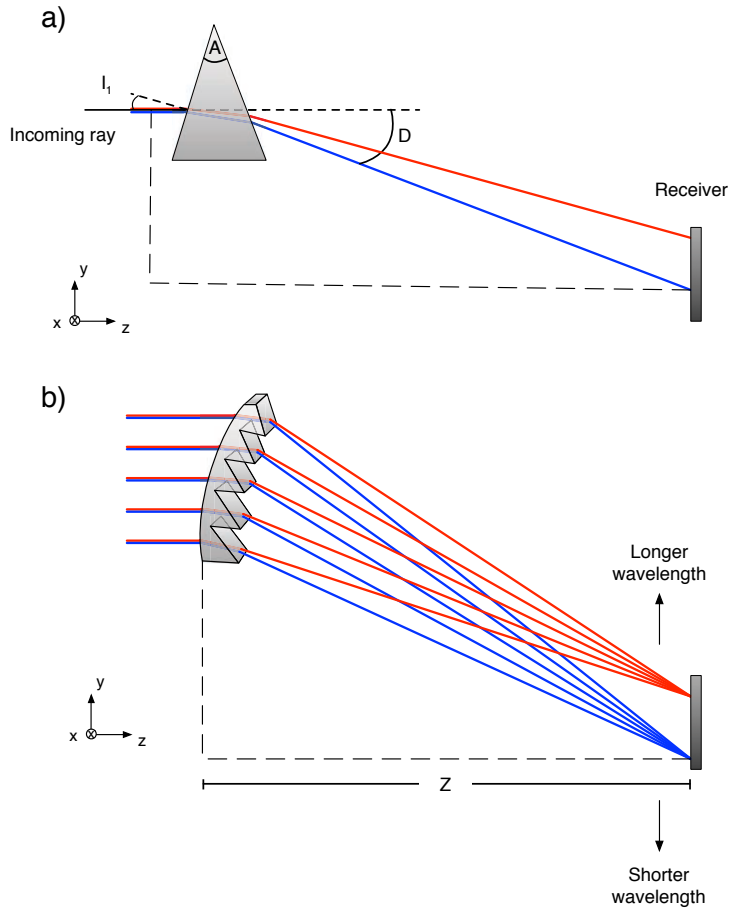

Fig. 1. a) Splitting of a single triangular prism. The two wavelengths depicted are separated in space at the focal plane according to the dispersive behaviour of the material as well as the geometrical parameters of the prism. b) Conceptual drawing of the prismatic spectral splitting concentrator presented in [24] with a detail of the beam splitting effect through part of the component. The figure shows that the element is able to deflect two reference wavelengths on two distinct points of the receiver.

The design described till now is two-dimensional. In [24] we extended the design of the element in the third dimension simply by extruding the two-dimensional contour in the direction perpendicular to the page plane (x direction). This resulted in a linear one-axis concentrator (denominated single element spectral splitter concentrator) whose design is illustrated in Figure 2a. The device was experimentally realized by mechanical machining and the spectrum obtained at the focal plane was analysed (Figure 2b). Although a good spectral separation was obtained along the y-direction, confirmed also by the spectral analysis[24], the light pattern obtained at the focal plane was big compared to the dimensions of an ordinary concentration photovoltaic cell (1x1 $cm^2$), particularly along the x-direction. This resulted in a low concentration factor and in a poor efficiency of the overall spectral splitting conversion system (data not shown). In addition the fabrication process ended up being expensive (around 1k$ per piece) given the dimensions of the device (approximately 7x11 cm).

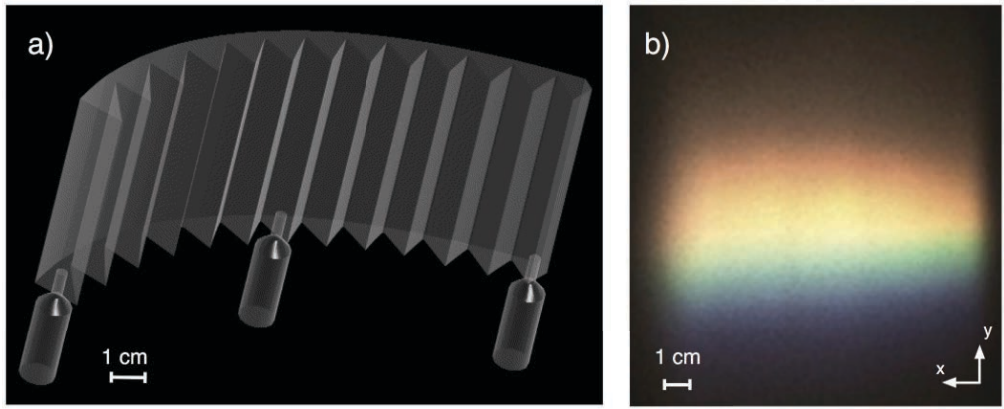

Fig. 2 a) Illustration of the single-element spectral splitter concentrator presented in [24]. It consists of a sequence of triangular prisms arranged along a curved line. It was designed in two-dimension and then extended in the third dimension simply by extruding the planar design. b) Picture of the spectrum obtained at the focal plane with the single-element spectral splitter concentrator. It shows clear separation within the visible part of the solar spectrum.

The main limitations of the design in [24] were in summary the ability to concentrate and split the light along a single dimension (y) and the high cost. Evolving from that design, we attempted to add concentration also along the x dimension and to lower the cost of the device by shrinking its size.
To address the first point we followed the procedure illustrated in Figure 3a where, conceptually, the x-extruded element was divided into an odd number of sections by planes perpendicular to the x-axis. While the central section was left unchanged, the other sections were modified by properly tilting each exit facet of a small angle u around a vertical axis parallel to y. This tilting allows introducing a beam deflection in the x-z plane, calculated for each section by using Snell's law to superimpose the section-generated image with the one of the central portion (Figure 3b).

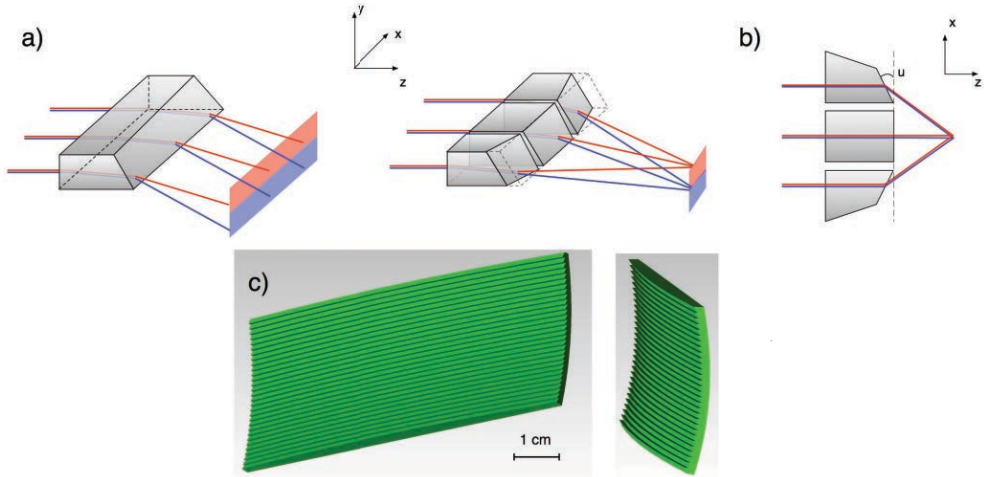

Fig. 3 a) Splitting of a single prism for the old (left) and new (right) design. The prism on the left splits and concentrates the colours along the y direction, while the one on the right concentrates also along the x direction. b) x-z section of the new design. The original element was divided in an odd number of sections and non-central portions were modified by tilting the exit facet of a small angle u. This allows for concentration in the x-direction c) Design of the point-focus spectral splitter concentrator. The element is composed of 30 trapezoidal prisms arranged in a curved line. Such configuration allows the element to spatially separate different wavelength bands of the solar spectrum while concentrating the light into a small spot on the focal plane.

As multiple possibilities exist for the choice of the tilt angle u, we performed this task preserving the continuity of the overall element facets to allow for manufacturability. Spectral splitting along the x direction is essentially non-existent as the facets angle in the x-z plane is extremely small with respect to that on the y-z plane. Still, computer based optimization was necessary to obtain a thin concentrated beam profile in the x direction. As a consequence, instead of a rectangular concentrated region, where splitting occurs only in the y direction, a slightly concave profile is obtained, without significant impact on the overall system performance.
In addition to that, we sensibly reduced the size of the element to allow for its fabrication by injection molding: this allowed us to increase the number of prisms while still cutting the cost of the element mold. This is an essential step in conceiving, from the very beginning, a design compatible with mass production constraints. Fig. 3c shows two illustrations of the final design of the element. The approximate area of the device is 7x3 cm$^2$, with an average thickness of 2 mm. Within its volume, 30 prisms are arranged.

## 3. Ray-tracing based optical analysis

In order to verify our design, we imported the model into a commercial ray-tracing tool (TracePro) and realistic optical simulations were performed to verify the predicted behaviour. A controlled divergence beam impinging on the concentrator was defined according to the angular distribution of the sunlight (max. divergence angle 4.8 mRad[25]). The parameters used for the simulations (material dispersion curve, number of rays, etc.) are reported in the supplementary material. The receiver is positioned 36 cm away from the axis of the optical element and 23 cm below with respect to the center of the optical element. Such distance between the spectral splitter and the receiver is required to achieve spectral separation. Shorter focal distances and therefore more compact systems can be obtained using steeper prisms at the expense of higher optical losses due to partial internal reflections at the exit facets. Figure 4a shows the results of ray-tracing simulations for two separate wavelengths, demonstrating how the optical element separates and concentrates the two different wavelengths. The figure shows that rays of different wavelengths are focused on different areas of the receiver, confirming the predicted behaviour. The same simulation can be extended to a greater number of wavelengths to determine the spreading along the solar spectrum. Figure 4b shows a two-dimensional map of the light intensity obtained at the receiver surface for different wavelengths (480, 532, 650, 1000 and 1500 nm). A separation of around 4.5 cm is obtained between the two extreme wavelengths, allowing to easily accommodate at least three photovoltaic cells to convert the light. Within the visible range, a 3-cm-long light pattern is obtained. With respect to the spectrum obtained with our previous optical element, the new design allows for higher geometrical concentration. Simulations show indeed that the optical element is capable of concentration ratios reaching up to 210x for a single wavelength against 140x for the old design (see supplementary material). It has to be noted however that, due to the splitting, the concentration is reduced with respect to the geometrical one and it becomes dependent on the considered wavelength region. For polychromatic light, considering a wavelength band between 710 and 1000 nm, the beam is deflected on a region of area 0.6x1.6 cm$^2$ (including beam divergence effects) with an effective concentration averaging 11x. The wavelength band from 450 to 710 nm is affected by a much larger change in refractive index of the material and is, therefore, spread out over a region of around 5 cm in length, giving an average concentration factor of 4.2x. Simulated results for the wavelengths between 1000 and 1500 nm provide a 0.7x1 cm$^2$ collection region with concentration around 20.5x. The results are summarized in Table I while a complete description of the calculations is reported in the supplementary material. The calculation is performed to determine the light concentration assuming that one PV receiver is used to convert each wavelength band. By increasing the number of bands, and therefore the number of PV cells, the concentration for each band will increase together with the efficiency of the overall converting system. It is to be noted that, depending on the dimension of the receivers, the illumination might be not uniform, thus affecting the efficiency of the system [26].

**Table 1. Geometrical concentration for different wavelength bands obtained on the receiver plane**

|  | 450-710 nm | 710-1000 nm | 1000-1500 nm |
|---|---|---|---|
| Geometrical concentration | 4.2 x | 11x | 20.5x |

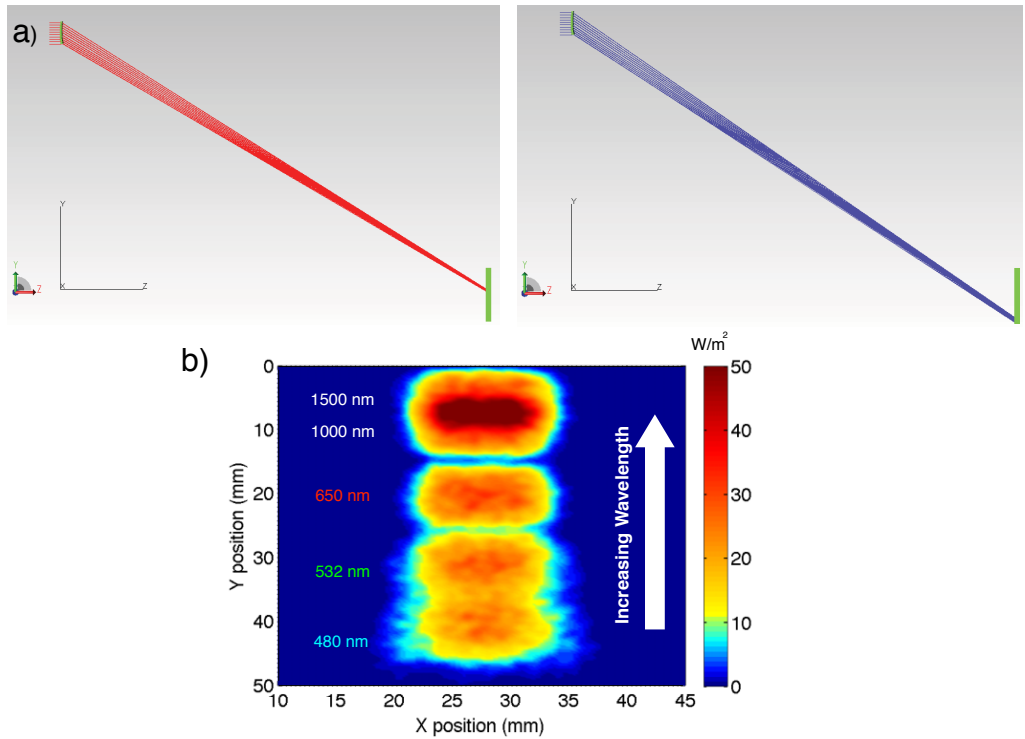

Fig. 4 a) Simulations of the optical element behaviour illuminated with collimated monochromatic light at two different wavelengths (480 and 650 nm). The figure shows that depending on the wavelength a different focal point on the receiver is obtained. b) Two-dimensional map of the light intensity obtained at the receiver surface. For this simulation 5 wavelengths were set (480, 532, 650, 1000 and 1500 nm). The intensity at the source was 10 W/m$^2$. The figure shows clear separation among different 'colours'.

## 4. Optical element realization and characterization

The optical element was fabricated by injection molding, a manufacturing process in which the material is fed into a heated barrel, mixed, and finally forced into a mold cavity, where it cools and hardens to the configuration of the cavity. For this technique, no mechanical processing of the element is needed: the facets flatness and prism apexes are limited only by the mold quality and the fluid dynamical specification of the material. In terms of quality, injection molded components can achieve, with post processing treatment, a RMS roughness as small as 15 nm[27, 28]. For this technique an initial investment (around 10 k$) is needed to fabricate the mold; however, once the mold is ready, the fabrication cost of each element decreases significantly with the number of pieces.

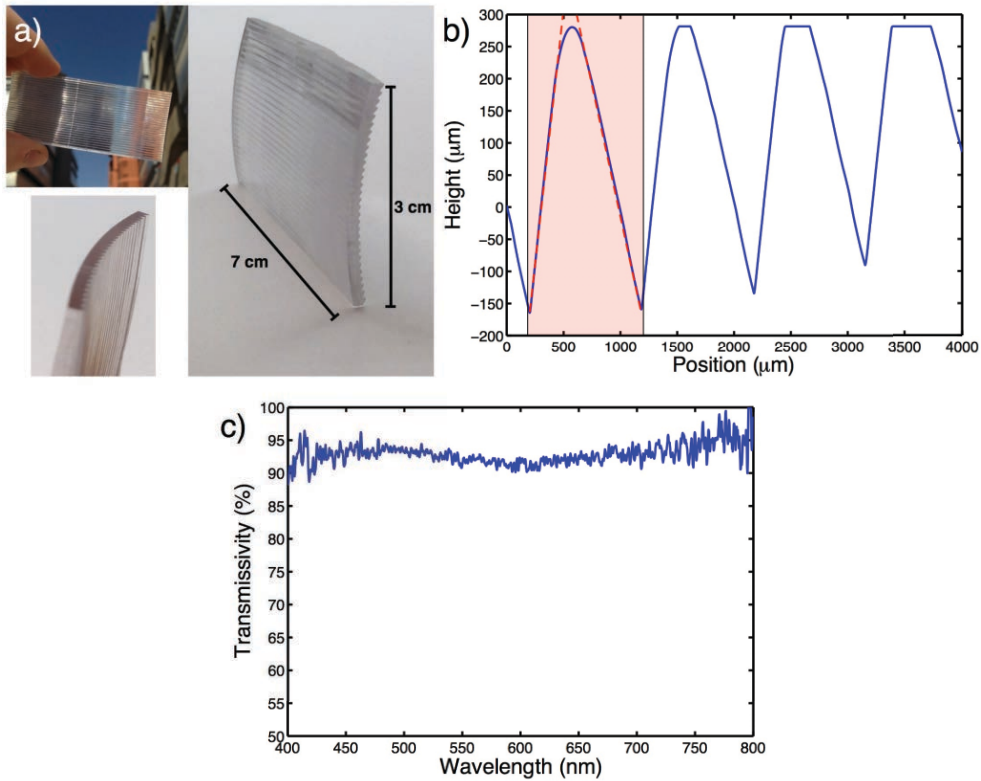

Fig. 5 a) Pictures of the optical element showing the actual dimensions of the device. b) Line profile of the textured surface of the optical element acquired with a profilometer. The shaded area of the graph shows also the comparison with the original profile of the device. c) Optical transmissivity of the element measured in the range 400-800 nm by using a spectrometer connected to an integrating sphere and illuminated with a polychromatic light source.

Figure 5a shows a set of pictures of the optical element. The textured surface of the optical element was scanned with a profilometer to verify its shape. Figure 5b shows a line profile. Within the shaded area, the measured profile is compared with the original design: a good matching is obtained, apart at the interface between two prisms where the discrepancy, in the form of rounding effect, is probably due to polymer retraction during the cooling phase and/or insufficient pressure during the injection[29]. Figure 5c shows the transmissivity of the optical element over the range 400-800 nm. The measurement was carried out using a multi-chromatic light source coupled with an integrating sphere feeding to an Ocean Optics USB4000 spectrometer. The measurement shows that an optical transmissivity above 90% is obtained in the range of analysis, essentially limited by the reflection at the entrance surface: such result reveals that the optical element has very low optical losses. Nevertheless, higher efficiencies can be achieved by the use of antireflection coatings on the plastic.

To extend the analysis also to the NIR and IR region, we report the transmissivity of PC in the supplementary material. Although PC is reported to have reliability and lifetime issues under sunlight exposure, UV degradation of PC can be inhibited by proper additive addition and its scratch resistance can also be improved by proper treatment[30, 31].

To test the element in real condition it was chosen to operate under real solar illumination. This approach is advisable because the correct operation of the optical element relies on the limited divergence of solar light[25] and on a spectral match of the source with the sun. The effective beam separation is strongly affected by the optical divergence of the light source, which in solar simulators can be an order of magnitude greater than the sun's divergence. An analysis of the performance of the optical device for different divergence angles of the light source is reported in the supplementary material.

Figure 6a shows the light pattern obtained at the focal plane. The colours are distinct, however the dimensions of the spectrum are not exactly equal to those predicted by the simulations. First, the spectrum is wider than expected and second, the colour dispersion is less pronounced than in the simulations (the visible light spreads along the y-axis for less than 2 cm, compared to a prediction of 3 cm). A reason for the first difference might be the divergence of solar light: the experiments were conducted in Abu Dhabi (United Arab Emirates), where studies performed on the direct solar beam divergence showed that the apparent dimension of the solar disks is significantly larger than expected due to humidity and high atmospheric dust levels[32, 33]. The less pronounced colour dispersion is in our opinion attributable to a slight difference in the dispersion curve of the employed PC with respect to the simulated curve, probably due to differences in the PC grade. To confirm this hypothesis we performed simulations whose results can be found in the supplementary material. In addition to that it can be noted that the contribution of some prisms falls outside the designed area, as can be appreciated in the bottom part of the Figure 6a where little 'rainbow-like' patterns can be seen. This is again probably due to inaccuracies in the manufacturing process of the optical element that in the end reduce the maximum concentration achievable. The colour separation was also confirmed by a spectral analysis with a moving probe (an optical fiber with high numerical aperture) feeding to a spectrometer. Figure 6b shows the light spectrum measured on three different spots at the center of the pattern. The fiber was mounted on a x-y stage and the data was acquired on three spots on the same x position, but on different y, as indicated by the markers in Fig. 6a. Each curve was then normalized with respect to their relative maximum to appreciate more easily the differences. The figure confirms the spectral separation, further validating the results shown in Figure 6a.

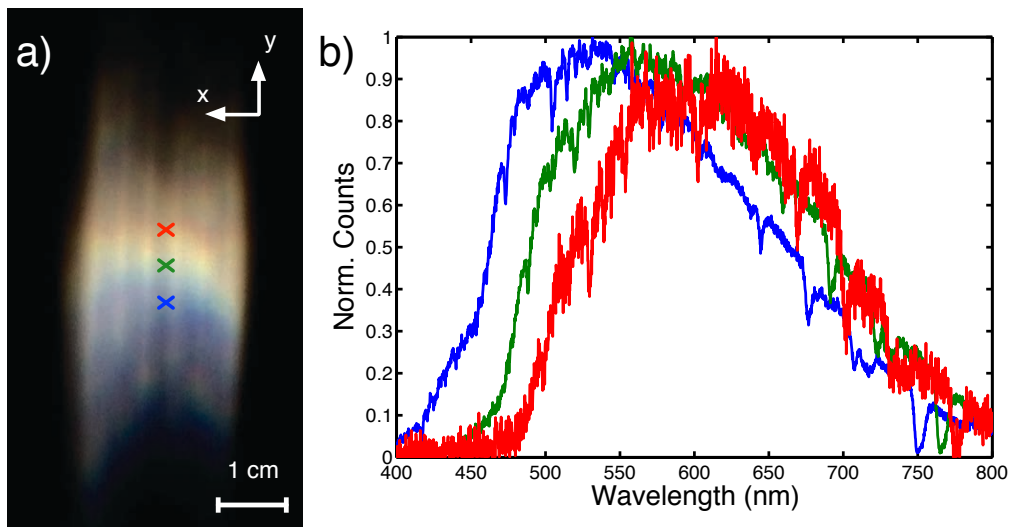

Fig. 6 a) Picture of the spectrum of the optical element obtained at the focal plane. The light pattern is limited in space in both directions and shows the different colours separated along the y direction. b) Spectral content of the light pattern measured on three different spots along the y direction. The measurements were carried out with a 0.3 NA optical fiber. The separation along the y-direction between contiguous measurements was 0.5 cm. The graph shows a clear shift of the spectrum peak from the blue to the red curve.

**5. Considerations about possible photovoltaic converting schemes**

So far we have demonstrated that the optical element described above is able to simultaneously separate different wavelength bands of the solar spectrum and to concentrate them in two directions. The spectral splitter represents however just half of the conversion system: in order to convert the sunlight, the optical element has to be indeed combined with a set of solar cells.

As previously introduced, the presented design aims to decouple the cells used for different spectral conversion, providing a platform where, contrary to MJ cells, no lattice matching constraints exist. This allows for a larger set of spectrally matched converters, possibly sourced among different technologies. Fig. 7 shows three selections of possible combinations of three 1x1 cm$^2$ cells positioned accordingly to the spectrum of our optical element, here replicated in scaled dimensions from the experimental results, and to their spectral response. In each selection, solar cells belonging to the same generation are shown (for a detailed description of each generation refer to [34]). Three receivers combinations are shown for each set; however, the number of receivers can be increased by optimizing their size. In the first set of absorbers, wafer-based materials are used to convert the sunlight. While such scheme allows for high conversion efficiencies (a theoretical study is reported in [35]), the high manufacturing costs of these materials is a deterrent for the development of spectral splitting systems. Nevertheless we performed simulations to determine the conversion efficiency, obtaining values around 30 %. Further details on the simulations can be found in the supplementary material. The second set is comprised of thin-film solar cells. $CuIn_{1-x}Ga_xSe_2$ is an interesting candidate for spectral splitting applications because of its composition-dependent band gap: its band gap varies continuously with x from about 1 eV (x=0) to about 1.7 eV (x=1)[36] and of recent development in bandgap tunable processing [37]. Therefore, short wavelengths can be converted by Ga rich $CuIn_{1-x}Ga_xSe_2$, while mid wavelengths can be converted by Ga poor $CuIn_{1-x}Ga_xSe_2$. A study of Ga poor $CuIn_{1-x}Ga_xSe_2$ solar cells under limited wavelength-range application can be found in [38, 39]. For long wavelengths, consideration could be given to $Hg_{1-y}Cd_yTe$ cells (band gap = 0.6 eV). With respect to the first scheme, thin-film cells allow reducing the cost of materials (thin film solar cells are generally in between one and two order of magnitudes thinner than wafer-based solar cells). However, their efficiencies are lower with respect to wafer-based solar cells and their behaviour has not been thoroughly studied under concentrated illumination. Finally, the last set of absorbers consists of colloidal quantum dots (CQD) and nano crystal (NC) solar cells. The ability to tune the band gap of CQD and NC by changing their size as well as their low manufacturing process costs make this generation of solar cells particularly promising for spectral splitting and MJ cells[40]. However, at the current state, further development is needed for this generation of solar cells to compete with the others. In any case, spectral splitting technology is inherently "cell agnostic", thus catalysing a broad competition between different promising cells technologies.

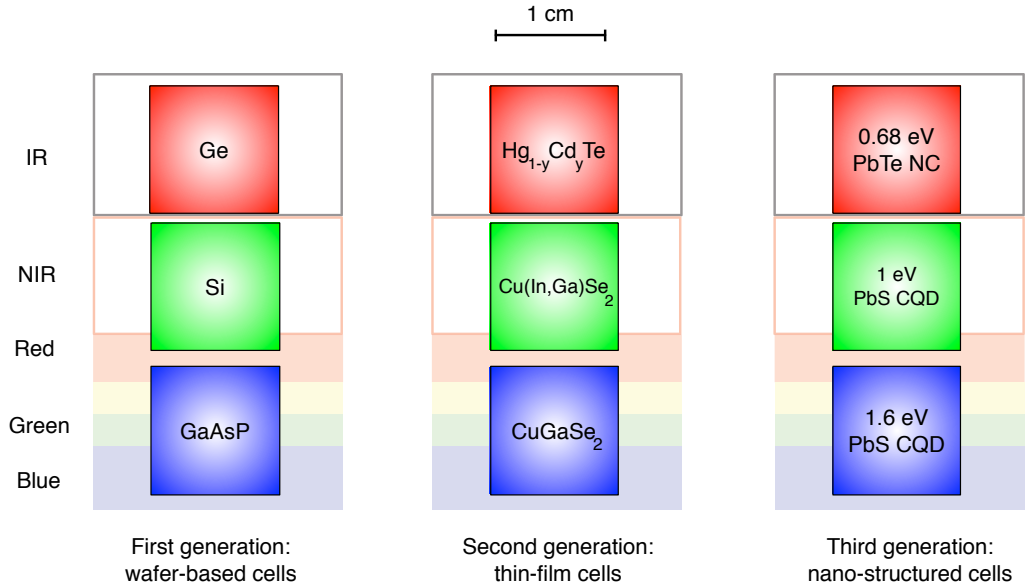

Fig. 7 Schematic of solar cells to be used in combination with the optical element. Three sets of receivers belonging to three different generations are shown. The dimensions of the cells (1x1 cm$^2$) are in scale with the light pattern obtained from the optical element.

## 6. Conclusions

In this paper we present a low-cost design of a dispersive optical element for the splitting and concentration of the solar light. The conceptual framework and mathematical building model used in designing the element, made of a sequence of dispersive trapezoidal prisms, is discussed and the design is verified with ray-tracing simulations. We show that the optical element is capable of concentrating the incoming light into a pattern whose size is compatible with the dimensions of concentrating photovoltaic cell. The device, intended from the very beginning taking into account industrial production constraints, is fabricated by injection molding and is characterized in terms of geometrical and optical properties. We show that the results of the characterization are in line with what theoretically predicted. Optimization in the fabrication of the element (using the correct grade of PC and optimizing temperature and pressure conditions) is required to further improve its performance. In conclusion our device represents a potential solution for economically viable spectral splitting photovoltaic applications. The experimental analysis of the complete converting system (optical element + solar cells) will be object of another publication.

## Acknowledgments

We would like to thank Mrs. Yamila Omar for her technical support and for proofreading the manuscript.

## References

[1]   D. S. Ginley and D. Cahen, *Fundamentals of materials for energy and environmental sustainability*: Cambridge university press, 2011.
[2]   J. Yang, E. Ziade, C. Maragliano, R. Crowder, X. Wang, M. Stefancich, M. Chiesa, A. K. Swan, and A. J. Schmidt, "Thermal conductance imaging of graphene contacts," *Journal of Applied Physics,* vol. 116, pp. -, 2014.
[3]   J. Yang, C. Maragliano, and A. J. Schmidt, "Thermal property microscopy with frequency domain thermoreflectance," *Review of Scientific Instruments,* vol. 84, pp. -, 2013.

[4]   W. Shockley and H. J. Queisser, "Detailed Balance Limit of Efficiency of p-n Junction Solar Cells," *Journal of Applied Physics,* vol. 32, pp. 510-519, 1961.

[5]   A. Polman and H. A. Atwater, "Photonic design principles for ultrahigh-efficiency photovoltaics," *Nature materials,* vol. 11, pp. 174-177, 2012.

[6]   H. Cotal, C. Fetzer, J. Boisvert, G. Kinsey, R. King, P. Hebert, H. Yoon, and N. Karam, "III–V multijunction solar cells for concentrating photovoltaics," *Energy & Environmental Science,* vol. 2, pp. 174-192, 2009.

[7]   F. Dimroth, "World record solar cell with 44.7% efficiency," Press Release, Freiburg, 23 Sep. 2013, No. 22/13, page 1.

[8]   M. Yamaguchi, T. Takamoto, K. Araki, and N. Ekins-Daukes, "Multi-junction III–V solar cells: current status and future potential," *Solar Energy,* vol. 79, pp. 78-85, 2005.

[9]   A. Braun, A. Vossier, E. A. Katz, N. J. Ekins-Daukes, and J. M. Gordon, "Multiple-bandgap vertical-junction architectures for ultra-efficient concentrator solar cells," *Energy & Environmental Science,* vol. 5, pp. 8523-8527, 2012.

[10]  V. Sorianello, L. Colace, C. Maragliano, D. Fulgoni, L. Nash, and G. Assanto, "Germanium-on-Glass solar cells: fabrication and characterization," *Optical Materials Express,* vol. 3, pp. 216-228, 2013.

[11]  M. Stefancich, C. Maragliano, M. Chiesa, S. Lilliu, M. Dahlem, and A. Silvernail, "Optofluidic approaches to stationary tracking optical concentrator systems," 2013, pp. 88340C-88340C-6.

[12]  A. Mojiri, R. Taylor, E. Thomsen, and G. Rosengarten, "Spectral beam splitting for efficient conversion of solar energy—A review," *Renewable and Sustainable Energy Reviews,* vol. 28, pp. 654-663, 2013.

[13]  A. Imenes and D. Mills, "Spectral beam splitting technology for increased conversion efficiency in solar concentrating systems: a review," *Solar energy materials and solar cells,* vol. 84, pp. 19-69, 2004.

[14]  R. K. Kostuk and G. Rosenberg, "Analysis and design of holographic solar concentrators," in *Solar Energy+ Applications*, 2008, pp. 70430I-70430I-8.

[15]  D. Zhang, M. Gordon, J. M. Russo, S. Vorndran, M. Escarra, H. Atwater, and R. K. Kostuk, "Reflection hologram solar spectrum-splitting filters," in *SPIE Solar Energy+ Technology*, 2012, pp. 846807-846807-10.

[16]  M. D. Escarra, S. Darbe, E. C. Warmann, and H. A. Atwater, "Spectrum-splitting photovoltaics: Holographic spectrum splitting in eight-junction, ultra-high efficiency module," 2013.

[17]  G. Kim, J. A. Dominguez-Caballero, H. Lee, D. J. Friedman, and R. Menon, "Increased Photovoltaic Power Output via Diffractive Spectrum Separation," *Physical Review Letters,* vol. 110, p. 123901, 2013.

[18]  Q. Huang, J. Wang, B. Quan, Q. Zhang, D. Zhang, D. Li, Q. Meng, L. Pan, Y. Wang, and G. Yang, "Design and fabrication of a diffractive optical element as a spectrum-splitting solar concentrator for lateral multijunction solar cells," *Applied optics,* vol. 52, pp. 2312-2319, 2013.

[19] G. Kim, J. A. Domínguez-Caballero, and R. Menon, "Design and analysis of multi-wavelength diffractive optics," *Optics express,* vol. 20, pp. 2814-2823, 2012.

[20] P. Wang, J. A. Dominguez‐Caballero, D. J. Friedman, and R. Menon, "A new class of multi‐bandgap high‐efficiency photovoltaics enabled by broadband diffractive optics," *Progress in Photovoltaics: Research and Applications,* 2014.

[21] J. D. McCambridge, M. A. Steiner, B. L. Unger, K. A. Emery, E. L. Christensen, M. W. Wanlass, A. L. Gray, L. Takacs, R. Buelow, and T. A. McCollum, "Compact spectrum splitting photovoltaic module with high efficiency," *Progress in Photovoltaics: Research and Applications,* vol. 19, pp. 352-360, 2011.

[22] B. Mitchell, G. Peharz, G. Siefer, M. Peters, T. Gandy, J. C. Goldschmidt, J. Benick, S. W. Glunz, A. W. Bett, and F. Dimroth, "Four‐junction spectral beam‐splitting photovoltaic receiver with high optical efficiency," *Progress in Photovoltaics: Research and Applications,* vol. 19, pp. 61-72, 2011.

[23] M. Peters, J. C. Goldschmidt, P. Löper, B. Groß, J. Üpping, F. Dimroth, R. B. Wehrspohn, and B. Bläsi, "Spectrally-selective photonic structures for PV applications," *Energies,* vol. 3, pp. 171-193, 2010.

[24] M. Stefancich, A. Zayan, M. Chiesa, S. Rampino, D. Roncati, L. Kimerling, and J. Michel, "Single element spectral splitting solar concentrator for multiple cells CPV system," *Opt. Express,* vol. 20, pp. 9004-9018, 2012.

[25] P. Bendt and A. Rabl, "Optical analysis of point focus parabolic radiation concentrators," *Applied optics,* vol. 20, pp. 674-683, 1981.

[26] H. Baig, K. C. Heasman, and T. K. Mallick, "Non-uniform illumination in concentrating solar cells," *Renewable and Sustainable Energy Reviews,* vol. 16, pp. 5890-5909, 2012.

[27] M. T. Gale, "Replication techniques for diffractive optical elements," *Microelectronic Engineering,* vol. 34, pp. 321-339, 1997.

[28] D. W. Sweeney and G. E. Sommargren, "Harmonic diffractive lenses," *Applied optics,* vol. 34, pp. 2469-2475, 1995.

[29] R. Chang and B. Tsaur, "Experimental and theoretical studies of shrinkage, warpage, and sink marks of crystalline polymer injection molded parts," *Polymer Engineering & Science,* vol. 35, pp. 1222-1230, 1995.

[30] D. C. Miller and S. R. Kurtz, "Durability of Fresnel lenses: a review specific to the concentrating photovoltaic application," *Solar Energy Materials and Solar Cells,* vol. 95, pp. 2037-2068, 2011.

[31] D. C. Miller, M. T. Muller, M. D. Kempe, K. Araki, C. E. Kennedy, and S. R. Kurtz, "Durability of polymeric encapsulation materials for concentrating photovoltaic systems," *Progress in Photovoltaics: Research and Applications,* vol. 21, pp. 631-651, 2013.

[32] M. Mokhtar, M. T. Ali, S. Bräuniger, A. Afshari, S. Sgouridis, P. Armstrong, and M. Chiesa, "Systematic comprehensive techno-economic

[33] P. R. Armstrong, R. Kalapatapu, and M. Chiesa, "Radiometers for measuring circumsolar profiles," ed: Google Patents, 2012.
[34] J. Jean, P. R. Brown, R. L. Jaffe, T. Buonassisi, and V. Bulovic, "Pathways for Solar Photovoltaics," *Energy & Environmental Science,* 2015.
[35] C. Maragliano, A. Zayan, and M. Stefancich, "Three-Dimensional Point-Focus Spectral Splitting Solar Concentrator System," *International Journal of Optics and Applications,* vol. 4, pp. 6-11, 2014.
[36] H. Luque, *Handbook of Photovoltaic Science and Engineering*, 2003.
[37] S. Rampino, N. Armani, F. Bissoli, M. Bronzoni, D. Calestani, M. Calicchio, N. Delmonte, E. Gilioli, E. Gombia, R. Mosca, L. Nasi, F. Pattini, A. Zappettini, and M. Mazzer, "15% efficient Cu(In,Ga)Se$_2$ solar cells obtained by low-temperature pulsed electron deposition," *Applied Physics Letters,* vol. 101, pp. 132107-13211, 2012.
[38] C. Maragliano, L. Colace, M. Chiesa, S. Rampino, and M. Stefancich, "Three-Dimensional Cu(InGa)Se$_2$ Photovoltaic Cells Simulations: Optimization for Limited-Range Wavelength Applications," *Photovoltaics, IEEE Journal of,* vol. PP, pp. 1-7, 2013.
[39] M. Stefancich, A. Zayan, M. Chiesa, S. Rampino, and C. Maragliano, "Single element point focus spectral splitting concentrator with CIGS multiple bandgap solar cells," in *SPIE Solar Energy+ Technology*, 2013, pp. 882108-882108-8.
[40] X. Wang, G. I. Koleilat, J. Tang, H. Liu, I. J. Kramer, R. Debnath, L. Brzozowski, D. A. R. Barkhouse, L. Levina, and S. Hoogland, "Tandem colloidal quantum dot solar cells employing a graded recombination layer," *Nature Photonics,* vol. 5, pp. 480-484, 2011.